\documentclass[12pt]{article}
\input epsf.sty
\usepackage{hyperref}

\usepackage{hyperref}
\usepackage{amssymb}
\usepackage{amsmath}
\usepackage[nosort]{cite}
\usepackage[normalem]{ulem}

\newcommand{\be}{\begin{equation}}
\newcommand{\ee}{\end{equation}}
\newcommand{\beq}{\begin{eqnarray}\displaystyle}
\newcommand{\eeq}{\end{eqnarray}}

\usepackage{amsmath}
\usepackage{amsfonts}
\usepackage{amssymb}
\usepackage{graphicx}

\def\sqr#1#2{{\vcenter{\vbox{\hrule height.#2pt
         \hbox{\vrule width.#2pt height#1pt \kern#1pt
            \vrule width.#2pt}
         \hrule height.#2pt}}}}

\usepackage{hyperref}
\hypersetup {
	colorlinks=true,
	linkcolor=blue,
	linktocpage=true,
	citecolor=blue,
	urlcolor=blue
}
\usepackage{geometry}

\usepackage{graphicx}

\usepackage{amsmath}
\usepackage{amsfonts}
\usepackage{amssymb}
\usepackage{graphicx}

\usepackage{setspace}
\numberwithin{equation}{section}

\begin{document}

\baselineskip=18pt

\begin{center}
{\Large \bf{A Puncture in the Euclidean Black Hole}}

\vspace{10mm}

\textit{Ram Brustein${}^{(1)}$,  Amit Giveon${}^{(2)}$, Nissan Itzhaki${}^{(3)}$ and Yoav Zigdon${}^{(1)}$ }
\break \break
(1)\ Department of Physics, Ben-Gurion University,
Beer-Sheva 8410501, Israel \\
(2) Racah Institute of Physics, The Hebrew University
Jerusalem, 91904, Israel \\
(3) School of Physics and Astronomy, Tel Aviv University, Ramat Aviv, 69978, Israel \\

\end{center}


\vspace{10mm}

\begin{abstract}

We consider the backreaction of the winding condensate on the cigar background.
We focus on the case of the $SL(2,\mathbb{R})_k/U(1)$ cigar associated with, e.g., the near-horizon limit of $k$ NS5 black-branes.
We solve  the equations of motion numerically in the large $k$ limit as a function of the amplitude,
$A$, of the winding mode at infinity. We find that there is a critical amplitude, $A_c=\exp(-\gamma/2)$, that admits a critical solution.
In string theory, the exact  $SL(2,\mathbb{R})_k/U(1)$ cigar CFT fixes completely the winding amplitude,
$A_s$, at infinity.  We find that in the large $k$ limit there is an exact agreement, $A_c=A_s$.
The critical solution is a cigar with a puncture at its tip; consequently, the black-hole entropy is carried entirely by the winding condensate.
We argue that, in the Lorentzian case, the information
escapes the black hole through this puncture.
\end{abstract}

\newpage

\vspace{10mm}

\section{Introduction}

The cigar geometry, which is the analytic continuation of the eternal black-hole solution, was proved to be a useful tool for the study of black-hole physics. For example, Gibbons and Hawking calculated~\cite{Gibbons:1976ue} thermodynamical properties of black holes~from the Euclidean action, an approach that is useful also in determining the quantum state associated with eternal black holes \cite{Hartle:1976tp}. More recently, a similar method to the Gibbons-Hawking one was used in~\cite{Penington:2019kki,Almheiri:2019qdq} to reveal the origin of the derivation in~\cite{Penington:2019npb,Almheiri:2019psf} of the Page curve~\cite{Page:1993wv}
(for a review, see~\cite{Almheiri:2020cfm}).

In string theory, the cigar geometry is accompanied by a winding  mode -- a string condensate that wraps the cigar.
At least far from the tip, the wave function of such a string is expected to take the form
\be\label{NG}
 \chi(\rho) \propto e^{-S_{NG}(\rho)}~,
\ee
where $S_{NG}(\rho)$  is the Nambu-Goto (NG) action for the worldsheet of a string
that wraps the cigar from its tip at the origin to a radial distance $\rho$ (see Fig.~1).

\begin{figure}[h]
	\centering
	\includegraphics[scale=0.5]{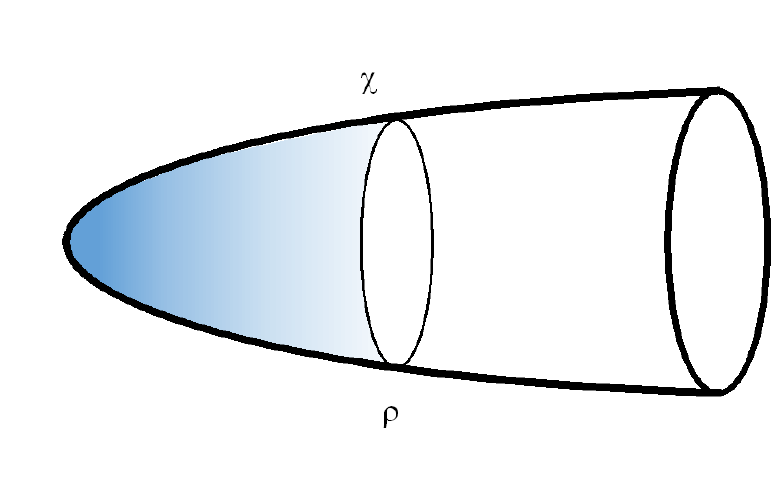}
	\caption{A string worldsheet wraps the cigar from the tip to some radial distance $\rho$. }
\end{figure}

It is widely accepted that $\chi$ plays a key role for small black holes near the Hagedorn temperature (see, e.g.,~\cite{Susskind:1993ws,Horowitz:1996nw,Kutasov:2005rr,Giveon:2005mi,Chen:2021emg,Chen:2021dsw}).
It was also claimed that this stringy mode is crucial for a microscopic understanding of large black holes
(see, e.g.,~\cite{Dabholkar:2001if,Giveon:2012kp,Mertens:2013zya,Giveon:2014hfa,Brustein:2021cza,Giveon:2021gsc}).
To explore this systematically, one needs to know the details of $\chi(\rho)$ and its coupling to the background.
Unfortunately, this is far from being the case for general cigar geometries and, in particular, to the Schwarzschild cigar.

The exact CFT description of the  $SL(2,\mathbb{R})_k/U(1)$ cigar, appearing in, e.g., the near-horizon limit of $k$ near-extremal NS5-branes in type II string theory~\cite{Maldacena:1997cg}, is extremely useful in this regard.
In particular, the Fateev, Zamolodchikov and Zamolodchikov (FZZ) duality~\cite{FZZ} makes~(\ref{NG}) precise~\cite{Giveon:2001up}.
Our goal here is to take advantage of this and to attempt to calculate the backreaction of $\chi$ on the $SL(2,\mathbb{R})_k/U(1)$ cigar geometry.

In the next section, we start with the Horowitz-Polchinski (HP)
equations of motion,~\cite{Horowitz:1997jc},
associated with the $SL(2,\mathbb{R})_k/U(1)$ cigar, and show that the order of the equations  can be reduced to accommodate the FZZ duality. We argue that this order reduction is crucial for the validity of the equations of motion at large $k$
(when the black hole is large compared to the string scale).
In section~3, we solve the reduced  equations of motion numerically at parametrically large $k$
as a function of the boundary condition of $\chi$ at infinity.
We find that a  critical solution appears {\it precisely} at the value determined by the boundary condition that is fixed in string theory via the FZZ correspondence \cite{FZZ,Giveon:2001up}. The critical solution involves a cigar with a puncture at its tip.
In section~4, we argue that the puncture at the tip removes the index obstruction, discussed recently in~\cite{Chen:2021dsw},
to smoothly connecting an HP classical solution with a black hole.
In section~5, we show that exactly at the critical value of the amplitude, the classical entropy carried by $\chi$ is the Bekenstein-Hawking entropy; the appearance of the puncture justifies it.
Section~6 is devoted to discussions, and in a couple of appendices, we present some technical details and derivations.

\section{From HP to FZZ}

In this section,  we start with the HP action~\cite{Horowitz:1997jc}
for the type II string on the $SL(2,\mathbb{R})_k/U(1)$ cigar and make contact with the FZZ correspondence~\cite{FZZ,Giveon:2001up}.

In this case, the HP effective action takes the form
\begin{eqnarray}
\label{HP}
I=\int d^2 x \sqrt{g}e^{-2\Phi}\left(-\frac{1}{2\kappa^2}(R-2\Lambda+4\partial^{\mu} \Phi \partial_{\mu} \Phi)+ \partial^{\mu}\chi \partial_{\mu}\chi^* + \frac{\beta^2 g_{\tau \tau}-\beta_H ^2}{4\pi^2 }\chi \chi^*\right),
\end{eqnarray}
where $\kappa^2 = 8\pi G_N$, $\Phi$ is the dilaton, $g_{\mu\nu}$ is the metric and $\chi,\chi^*$ are the $\pm 1$ winding modes.
We use units in which $\alpha'=1$, and so $\beta=2\pi\sqrt{k}, \beta_H = 2\pi \sqrt{2}$, $\Lambda=-\frac{2}{k}$,
and discuss the parametrically small curvature case, $k\gg 1$.

Our main goal is to explore solutions to the equations of motion  such that the 2D geometry is given by
\be\label{ansatz}
ds^2=h^2(\rho)d\tau^2+d\rho^2~,
\ee
where $\rho$ is the radial direction, $\tau$ is the thermal circle,
$\tau\sim\tau+\beta$,
and the dilaton and winding mode are functions of the radial coordinate, $\Phi(\rho)$ and $\chi(\rho)$, respectively.

Normally, the HP action is used near the Hagedorn transition,
\be
\label{con}
\frac{\beta-\beta_H}{\beta_H}\ll 1~,
\ee
in which case $\chi$ is light everywhere. Here, however, we are interested in the large $k$ limit. In this case,
\be\label{cond}
\beta\gg \beta_H~,
\ee
and condition (\ref{con}) is not satisfied.
This leads to the following apparent problems. The reason to focus on $\chi,\chi^*$  and ignore higher winding modes when (\ref{con}) is satisfied is that in this case the $w=\pm 1$ modes are light, while modes with $|w|>1$ have mass at least of order of the string mass scale. But, in the regime (\ref{cond}), the local mass of the $w=\pm 1$ modes is large at infinity.
In such a situation, one would expect, naively, higher winding modes and higher derivative terms to be important.

At least in the case of the $SL(2,\mathbb{R})_k/U(1)$ cigar,
we have a reason to expect that the situation is, in fact, much  better: The FZZ correspondence \cite{FZZ,Giveon:2001up},
which implies that the winding condensate $\chi(\rho)$ is not independent of the metric.
In fact, it is correlated with the zero mode of the metric that shifts the location of the tip of the cigar.
We can think of the winding mode and the graviton-dilaton zero mode as two semi-classical tails corresponding
to the same state in the CFT.
One way to see this is to note~\cite{Giveon:2016dxe} that, regardless of $k$, in the father $AdS_3$ CFT they amount to isomorphic representations
of affine $SL(2,\mathbb{R})$~\cite{Maldacena:2000hw}.
So, also for $k\gg 1$, the zero modes of the graviton, that are states with $w=0$,
are identical to states with $w=\pm 1$ (and not to states with $|w|>1$).
This explains why even when $k$ is large, states with $w=\pm 1$ should be included in the
low-energy effective action, despite the fact that the mass of $\chi$ is large at infinity
-- it is the $|w|=1$ zero mode that accompanies the $w=0$ zero mode associated with shifting the tip of the cigar.

What this does not explain is why higher derivatives
can be neglected.  A partial explanation for this is the following.
The HP action does not reflect these FZZ features explicitly, however, the order of the HP equations of motion can be reduced to account for (most of) the FZZ correspondence input.
In particular,
the fact that $\chi(\rho)$ is not independent of the metric implies that it should not have an independent kinetic term.
The way that we address this issue is the following: As expected, the equation of motion for $\chi(\rho)$ is second order.
However, there is a subset of solutions that satisfy first order equations for $\chi$,
which fit neatly with expectations from the FZZ correspondence.
In fact, as we shall see, all the equations of motion can be reduced to first order.
We view this as an indication that higher derivatives can be neglected.

To recapitulate, while the use of the HP equations of motion is not a priori valid for $k\gg1$, the use of their reduction, that takes account of the FZZ duality, is a justified  starting point.

We thus begin with the equations of motion resulting from the HP action (\ref{HP}).
After the rescaling
\be
h\to \sqrt{k} h~,~ \chi \to \kappa \chi~,
\ee
and straightforward calculations (see Appendix A), the equations of motion take the form:
\beq
\label{bad}
h\left(\frac{\Phi'}{h}\right)' & = & (\chi')^2+h^2\chi^2~,\nonumber \\
h\chi ''+h'\chi ' -2 h\chi ' \Phi'  &=&(h^2 - 2) h \chi~, \\
\frac{2}{k}+2\Phi''-2(\Phi')^2 &=& (\chi')^2 + (3h^2-2)\chi^2~. \nonumber
\eeq
Ultimately, we wish to study  solutions to the equations in~(\ref{bad}), including the full backreaction of $\chi$ on the dilaton and metric.
But, before we do that, let us recall some known properties of these equations when  $\chi$ is treated as a small perturbation.

Setting $\chi=0$, we find the cigar geometry,
\begin{eqnarray}\label{BC}
h(\rho)=\sqrt{k}\tanh\left(\frac{\rho}{\sqrt{k}}\right)~,  ~~~~~
e^{2\Phi}=\frac{e^{2\Phi_0}}{\cosh^2\left(\frac{\rho}{\sqrt{k}}\right)}~.
\end{eqnarray}
Treating $\chi$ as a perturbation, without including its backreaction on the geometry, the equation of motion  for $\chi$
implies~\cite{Giveon:2013ica} that
\begin{equation}\label{coshk}
\chi(\rho)=\frac{A}{\cosh^k\left(\frac{\rho}{\sqrt{k}}\right)}~,
\end{equation}
which interestingly enough exactly agrees with (\ref{NG}) for any $\rho$ \cite{Giveon:2019gfk}.
This can be viewed as a first indication that things work better than naively expected, as we shall see in the rest of this section.

For $\rho\gg \sqrt{k}$, we have
\be\label{twokc}
\chi= 2^k A e^{-\sqrt{k} \rho}~,~~~~\frac{h(\rho)}{\sqrt{k}}=1-2e^{-2\frac{\rho}{\sqrt{k}}}~,~~~~\Phi(\rho)=\Phi_0-\frac{\rho}{\sqrt{k}}~.
\ee
For $k\gg 1$, we can also consider $\rho\ll \sqrt{k}$, in which case
\be
\label{chGauss}
\chi=A e^{-\frac{\rho^2}{2}}~,
\ee
and the background is $\mathbb{R}^2$ with a constant dilaton,
\be
~~~~h(\rho)=\rho~,~~~~\Phi(\rho)=\Phi_0~.
\ee
These simple observations~\cite{Giveon:2012kp,Giveon:2013ica} will play an important role in what follows.

Next, we turn to discuss solutions for which the backreaction of the winding modes on the geometry is taken into account. As discussed above, the FZZ duality implies that the second order equation of $\chi$ should be replaced by a first order equation. Since the equations in~(\ref{bad}) are non-linear coupled equations, it is hard to imagine that a replacement of this kind is possible. But it is. One can show (see Appendix A) that any solution of the following set of equations,
\beq
\label{goku}
h' &=& h \Phi' +1 ~, \nonumber \\
\Phi' &=& -h\left(\chi^2+\frac{1}{k} \right)~,  \\
\chi ' &=& -h \chi~,\nonumber
\eeq
also solves the equations in~(\ref{bad}).
In complete accord with the FZZ duality, $\chi$ is not a dynamical field in the equations in~(\ref{goku});
rather, it is fixed by the metric.
Furthermore, the last equation of (\ref{goku}) implies that (\ref{NG})
holds everywhere also when the full backreaction of the winding mode is included.
Note that the equations of motion for $h$ and $\Phi$ are first order too.
This reflects the fact that there are no propagating modes in the 2D gravity-dilaton system.

In the rest of the paper,
we will use the FZZ reduced-order equations of motion (\ref{goku}) rather than the HP ones (\ref{bad}).

In string theory, the value of the amplitude, $A$, is fixed.
This is correct for any cigar geometry. However, since the calculation of $A$ involves non-perturbative corrections in $\alpha'$,
it can be calculated only when there is an exact CFT description of the relevant cigar.
This is the case in the $SL(2,\mathbb{R})_k/U(1)$ cigar.
The way this comes about is that the properties of the
coset CFT relate $\chi$ with the cigar background.
In particular, $A$ is entangled with the asymptotic behavior of $h$ and $\Phi$ in~(\ref{twokc}),~\cite{Giveon:2001up}.
In Appendix B, we review this and show that for large $k$,
\be\label{cfzz}
A_{s}=e^{-\gamma/2}=0.749306001...~,
\ee
where $\gamma$ is the Euler-Mascheroni constant.

However, (\ref{goku}) does not fix the amplitude $A$ in~(\ref{coshk},\ref{chGauss}).
Our approach in the next section is to solve the equations in~(\ref{goku}) numerically for various values of $A$.
We find that, within the numerical accuracy of our analysis, exactly at $A_c=A_s$, a critical  behavior appears
-- {\it there is a puncture at the tip of the cigar}.

\section{The puncture}

In this section, we present some properties of the solutions of (\ref{goku}).
First, by discussing the numerical solutions of the equations in~(\ref{goku}),
and then by considering an analytical solution that is valid near the puncture of the cigar.

\subsection{Numerical solutions}

Here, we describe the numerical solutions of (\ref{goku}) for $k\to \infty$.
The boundary conditions are fixed at $\rho=\tilde{\rho}$, with
$
1\ll \tilde{\rho} \ll \sqrt{k}~,
$ in the following way:
\be
\chi(\tilde{\rho}) = Ae^{-\frac{1}{2}\tilde{\rho}^2}~, ~~~~h(\tilde{\rho})=\tilde{\rho}~,~~~\Phi'(\tilde{\rho})=0~.
\ee
In the numerical analysis, we have often set the boundary conditions at $\tilde{\rho}=5$.
This boundary condition neglects backreaction of the winding mode and so it induces some uncertainty of the order of
\be
\chi (\tilde\rho)^2\sim A^2\times \exp(-5^2)\sim A^2\times10^{-11}~.
\ee
As we shall see, $A_c\sim 1$, so there is a built-in uncertainty of the order of $10^{-11}$ in our numerical solutions.

Depending on the value of $A$, we find that there are two classes of solutions.
In one class, $h(\rho)$ vanishes at some $\rho_0$ and $h'(\rho_0)=1$, namely, the geometry has a tip. As the amplitude approaches, from below, a critical amplitude, $A_c$, we find that $\rho_0$ becomes more and more negative, and a narrow neck develops.
In the second class, $h'(\rho_0) $ vanishes while $h(\rho_0)>0$, that is, in this class, the geometry has a hole.
As we approach $A_c$ from above, the size of the hole becomes smaller and smaller.
Moreover, in the second class, $h$, $\Phi$ and $\chi$ diverge at some $\rho<\rho_0$, which is not far (in string units) from $\rho_0$.
Examples of numerical solutions of $h(\rho)$, for several values of $A$, appear in Fig. 2. An example of the numerical solutions of
$h$, $\Phi'$ and $\chi$ for a value of $A$ which is slightly below $A_c$ is presented in Fig. 3.

There are various critical exponents that appear in the transition between the two classes; we describe some of them in Appendix C. Here, we focus  on the critical solution.  At the critical value  $A_c$, that separates the two classes, the size of the hole vanishes and we expect, based on the asymptotic analytic solution described in the next subsection, that $\rho_0\to -\infty$ (see Fig.~2). Namely, there is a puncture at the tip of the cigar.  The value of $A_c$  is
\be\label{critical}
A_c= e^{-\gamma/2},
\ee
within the numerical accuracy of our analysis.

The reason we focus  only on $A_c$ is that this is the case that is relevant in string theory,
since $A_c=A_s$ (see appendix B). Near the puncture, the critical solution can be described analytically.
This is done in the next subsection.

\subsection{Analytical description}

\begin{figure}\label{ht}
	\centering
	\includegraphics[scale=0.6]{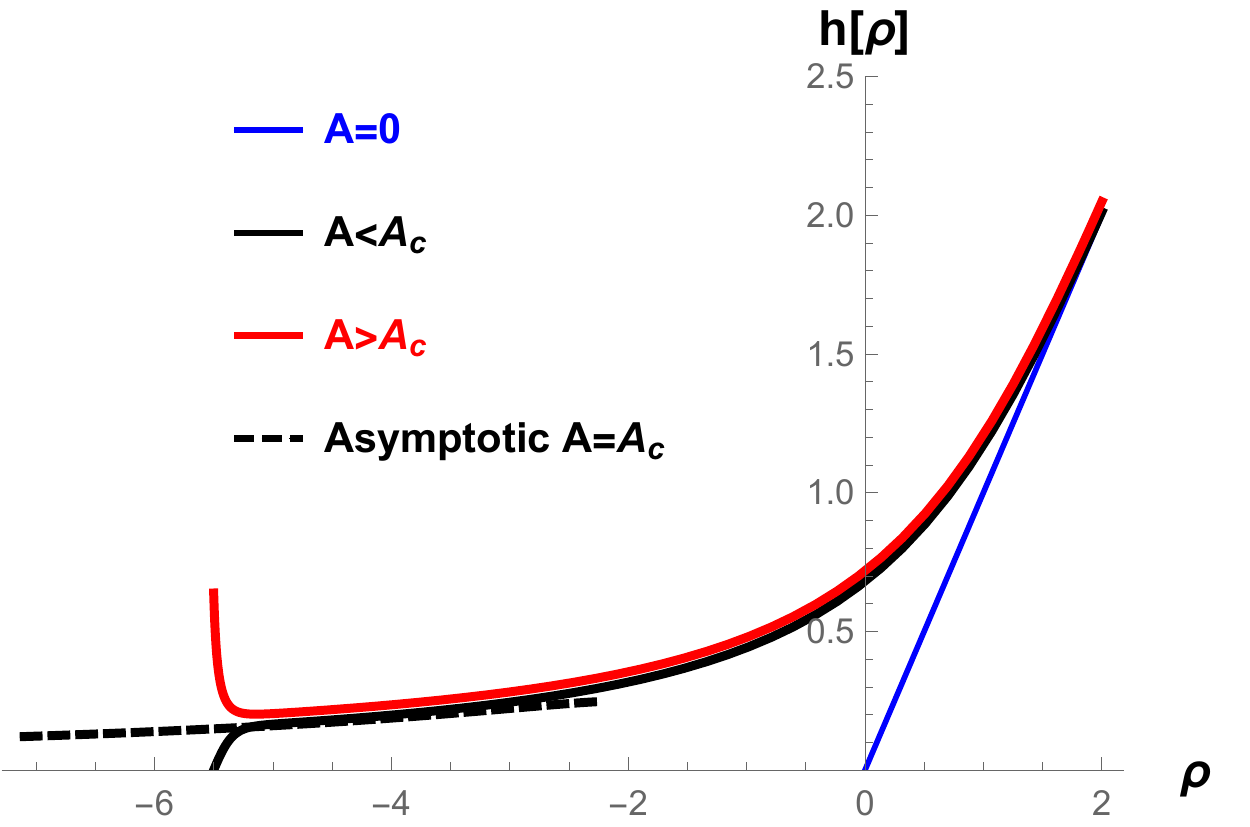}
	\caption{Shown are numerical solutions of $h(\rho)$, in the $k\to\infty$ limit,  for different values of $A$ (solid) and the analytical asymptotic solution from (\ref{h-}) (dashed).
The black curve is the plot of a solution for $A$ slightly below $A_c$.
The red curve is a solution for $A$ slightly above $A_c$; it is shifted up a bit to make it visible.
Below $A_c$, the geometry has a tip. Above $A_c$, it has a hole followed by a divergence.
The critical solution, for $A=A_c$, when both curves coincide and which connects smoothly to the asymptotic solution,
is the punctured Euclidean black hole. In the plot, both curves
have an amplitude $A$ that deviates from $e^{-\gamma/2}$ in~(\ref{cfzz}) only in the $11$th digit after the decimal point.}
\end{figure}

For $-\rho\gg 1$, the critical numerical solution is well approximated by
\begin{equation}
\label{h-}
h(\rho)=-\frac{1}{\rho}- \frac{1}{\rho^2}+\cdots~,
\end{equation}
\begin{equation}
\label{Phi-}
\Phi' (\rho) = \rho-1 +\cdots~,
\end{equation}
\begin{equation}
\label{chi-}
\chi(\rho) = -\rho+1+\cdots~.
\end{equation}
Indeed, one can verify that (\ref{h-})--(\ref{chi-})  solve (\ref{goku}) in the limit $\rho\to -\infty$.

From (\ref{h-}), whose plot is shown in Fig.~2, it is clearly seen that the asymptotic behavior of $h(\rho)$ at large negative $\rho$
is compatible with the numerical solutions. It is also obvious from (\ref{h-}) that both $h$ and $h'$ vanish when $\rho\to -\infty$.
The dilaton grows quadratically there, so formally it is a strong-coupling region, where quantum corrections are important.
However, the apparent need for adding more terms to the  action is perhaps less pressing than it might appear to be:
Taking $\Phi_0 \to -\infty$, the region where strong coupling appears is pushed to infinity,
and as we demonstrate in section 5, the entropy associated with the winding mode
comes from the region where the string coupling constant is small.

Since both $h$ and $h'$ vanish when $\rho\to-\infty$, this solution describes a tip with a puncture.
Near the puncture it is natural to apply a T-duality (in the angular direction).
The duality takes $h\to 1/h$, and so we get back $\mathbb{R}^2$, albeit with a non-trivial dilaton and an angular momentum mode that is the T-dual of $\chi$.

Note that around $\rho=0$, the critical solution has a curvature of order $1$ in string units.
The facts that $A_c=A_s$ and that, as discussed in the next section, $S_{BH}=S_W$,
suggest that the $\alpha'$ corrections do not enter in these calculations.
This is in accord with our argument in the previous section that higher derivatives can be ignored.
\begin{figure}[h]
	\centering
	\includegraphics[scale=0.6]{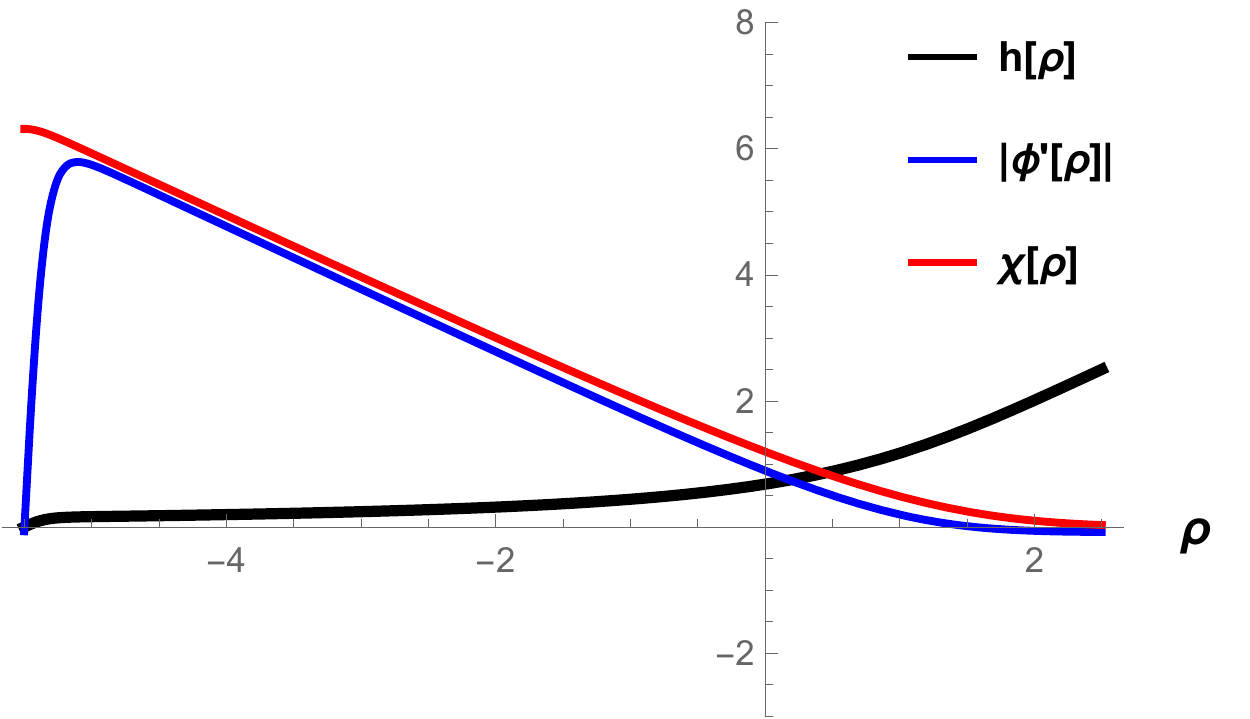}
	\caption{Solutions of $h(\rho), -\Phi'(\rho)$
 and $\chi(\rho)$, for the $A<A_c$ solution shown in Fig.~2. }
\end{figure}

\section{Witten index and the elliptic genus}

We argued that for large $k$, a puncture appears at the tip of the cigar in string theory. If this is the case for any $k$, then the Witten index obstruction to smoothly connecting an HP solution with a black hole in type II string theory, discussed recently in \cite{Chen:2021dsw}, is evaded in our case. Concretely, in the  $SL(2,\mathbb{R})_k/U(1)$ cigar, $k$ is the parameter that connects the HP solution with the black hole solution. For large $k$, we have the Euclidean black hole with Witten index 1, and for small $k$, we have a cylinder associated with the HP solution with Witten index 0.  The transition takes place at $k=1$ \cite{Giveon:2005mi}.

We argued that for large $k$ there is a puncture at the tip, so the topology is of a cylinder, just like the HP solution. This implies that there is no index obstruction in connecting smoothly HP with the black hole solution. Moreover, this suggests that  even for large $k$ the black hole  should be described at the microscopic level by the Wick rotation of the winding modes; more on that in the discussion section.

At first, the statement that the Witten index vanishes in the $SL(2,\mathbb{R})_k/U(1)$ CFT  seems to be in contradiction with known results. In a   beautiful work,~\cite{Troost:2010ud}, Troost calculated the elliptic genus associated with the $SL(2,\mathbb{R})_k/U(1)$ CFT and, in particular, showed that the Witten index is $1$, in agreement with semi-classical expectations.

The curious sum rule of~\cite{Giveon:2013hsa} (see also~\cite{Sugawara:2013hma}), however, seems to support a puncture at the tip.
Namely, a cylinder topology. The sum rule states that the sum of the following three elliptic genera vanishes. The first is the $SL(2,\mathbb{R})_k/U(1)$ cigar CFT geometry. The second is the  analytic continuation of the region  between the horizon and the singularity,
which gives an $SU(2)_{-k}/U(1)$ coset CFT (whose geometry is the bell). The third is the region beyond the singularity,
which gives a $Z_k$ orbifold of the $SL(2,\mathbb{R})_k/U(1)$ CFT (the T-dual of the trumpet).

The fact that their sum vanishes suggests that there is a sense in which the three regions, which semi-classically are distinct,
are secretly glued to form a background with the topology of a cylinder.
From this point of view, the results presented here imply that the glue that connects the three regions is the winding condensate.

\section{Winding entropy and black hole entropy}

In this section, we discuss the classical entropy carried by the winding mode, $S_W$.
We show that exactly for the stringy amplitude, $A_s$ in~(\ref{cfzz}),
\be
S_W=S_{BH}~,
\ee
where $S_{BH} =2\pi e^{-2\Phi_0}$ is the black-hole entropy (see, e.g.,~\cite{Giveon:2005mi}).
Additionally, we present some comments concerning the entropy carried by $\chi$ and its normalization.

The entropy carried by the winding condensates
can be expressed in terms of the action $I$ in (\ref{HP}),
\begin{equation}
S = (\beta \partial_{\beta}-1)I~,
\end{equation}
when there is a puncture at the tip of the cigar.
The calculation can proceed in two ways: (i) Calculate the $\beta$-derivative of the integrand of $I$ and then set the fields on-shell and (ii) Calculate the on-shell action  and then take the $\beta$-derivative; both yield the same answer.

Starting with the HP action in $(d+1)$-dimensions and taking a $\beta$ derivative on the integrand before setting the fields on-shell implies that the entropy is given by \cite{Chen:2021emg,Brustein:2021cza,Giveon:2021gsc,Chen:2021dsw}:
\begin{equation}\label{qp}
S_W = \frac{2\beta^3}{(2\pi)^2}\int  \sqrt{g} e^{-2\Phi} g_{\tau \tau} |\chi|^2 d^d x~.
\end{equation}
The entropy  $S_W$ is carried entirely by the winding mode $\chi$.
In our 2D case, (\ref{HP}),
\begin{equation}\label{SW}
S_W = 4\pi \int e^{-2\Phi(\rho)} h^3 (\rho) \chi^2 (\rho)d\rho~.
\end{equation}
Now, the $g_{\tau\tau}$ equation of motion (see Appendix A),
\begin{equation}\label{tteom}
h''-2h' \Phi' = 2h^3 \chi^2~,
\end{equation}
allows to express the integrand in (\ref{SW}) as a total derivative:
Plugging (\ref{tteom}) into (\ref{SW}) yields
\begin{equation}
S_W = 2\pi \int \left(e^{-2\Phi(\rho)} h'(\rho)\right)' d\rho~.
\end{equation}
This integral is given by the boundary terms: For $\rho\to \infty$, the backreaction of $\chi$ vanishes and the background takes the form (\ref{BC}), so the value of the boundary term at infinity is  $2\pi e^{-2\Phi_0}$. At $\rho\to -\infty$, we have
$h'(\rho)$ and $e^{-2\Phi(\rho)} \to 0$, so the boundary term there vanishes. Therefore,
\begin{equation}
S_W = 2\pi e^{-2\Phi_0}~,
\end{equation}
which is exactly equal to the Bekenstein-Hawking entropy of the black hole.

As we show next, calculating the on-shell action and then taking the $\beta$-derivative yields the same result.
Since the dilaton equation is satisfied, the on-shell action can be expressed in terms of a boundary term~\cite{Chen:2021dsw},
\begin{equation}
I = -\frac{1}{\kappa^2} \int_{\partial M} d^{D-1} y ~\partial_n \left(\sqrt{g_{ind}} e^{-2\Phi}\right)~,
\end{equation}
where $\partial_n$ denotes the normal derivative and $g_{ind}$ denotes the induced metric on the boundary.
In our case, the $\rho \to \infty$ boundary term is identical to the boundary term for a black hole with the same asymptotics.
The contribution from the boundary $\rho \to -\infty$ vanishes, due to the fact that both $h$ and $e^{-2\Phi}$ vanish there,
as can be seen from  eqs.~(\ref{h-}) and (\ref{Phi-}).
Thus, the action of the critical punctured cigar solution is the same as the cigar action,
so their free energies are equal and so are their entropies.

Additionally, we point out that  the winding modes action vanishes on-shell for the backreacted solution:
By integrating the kinetic term of $\chi$ by parts and using the $\chi$ equation of motion, which is in the middle of (\ref{bad}), we get
\begin{equation}\label{IW0}
I_W = 2\pi  \int d\rho ~ h e^{-2\Phi} \left[(\chi')^2 + (h^2-2) \chi^2\right]=2\pi \left( e^{-2\Phi} h \chi \chi' \right)\biggl|_{\rho\to -\infty}^{\rho\to +\infty}=0~,
\end{equation}
where the boundary terms vanish for both $\rho\to \infty$ and $\rho\to -\infty$.
We thus confirmed that $\chi$ remains a {\it zero mode} of the effective theory also after backreaction is taken into account.

Finally, we can relate the norm of $\chi$, defined by
\begin{equation}\label{norm}
N^2_\chi \equiv  4\pi \int d\rho~ e^{-2\Phi}~ h \chi^2~,
\end{equation}
to the entropy. The relation $\chi' = -h \chi$ of (\ref{goku}) and the zero-mode action in (\ref{IW0}) imply
\begin{equation}
 2\pi \int d\rho ~ h e^{-2\Phi} \left(2h^2-2\right) \chi^2=0~.
\end{equation}
From eqs. (\ref{norm}) and (\ref{SW}), it follows that
\begin{equation}
N^2_\chi = S_W = S_{BH}~,
\end{equation}
in harmony with the fact that the black-hole entropy is carried by the winding zero mode $\chi$.

One can 
calculate $S_W$ also for $A\neq A_s$, using~(\ref{SW}).
In Fig.~4, we plot the ratio of the entropy carried by the winding condensate to the black hole entropy  as a function of $A/A_c\leq 1$. As can be seen from the figure, $S_W=S_{BH}$ only for $A=\exp(-\gamma/2)$,
implying that the critical solution possesses the entire black hole entropy exactly when a puncture is formed.

\begin{figure}[h]
	\centering
	\includegraphics[scale=0.5]{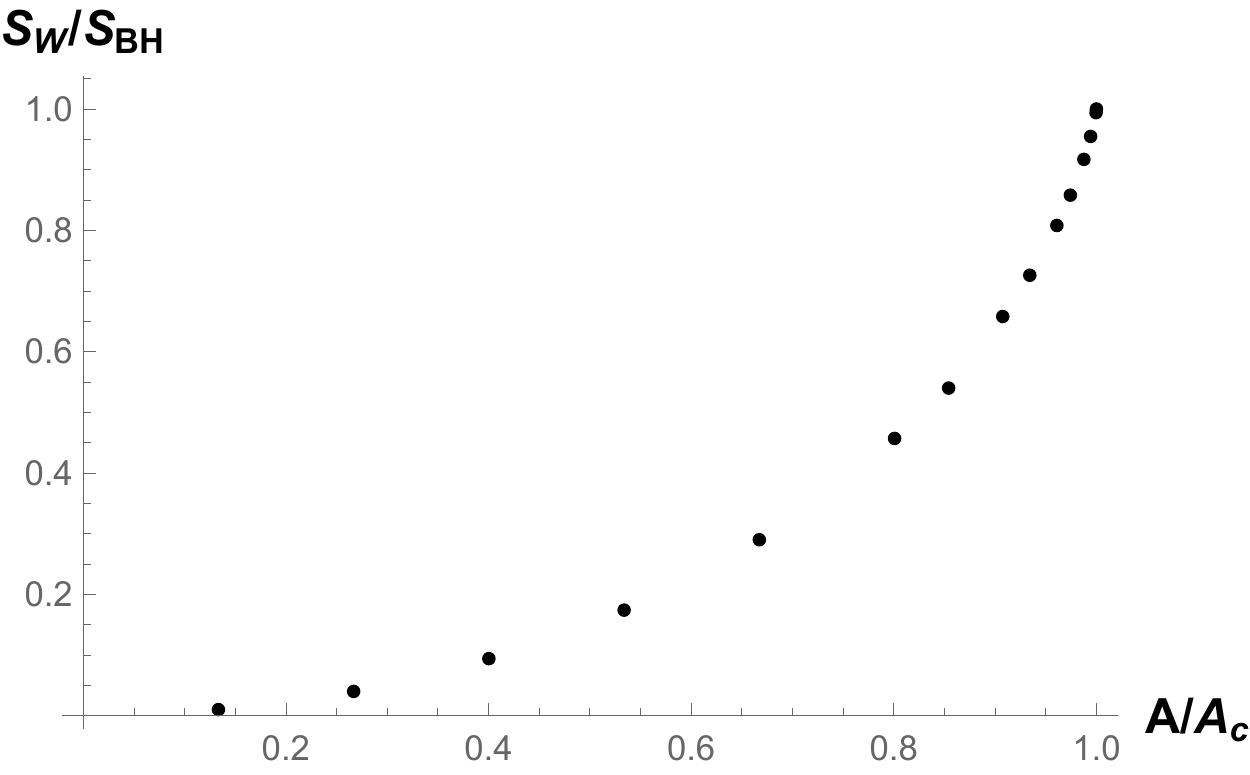}
	\caption{Shown is the ratio of  entropy of the winding-mode condensate to the black hole entropy for different values of $A$. The ratio approaches unity as $A$ approaches $A_c$ from below.}
\end{figure}

Figure 5 depicts the ratio of the entropy density of the winding mode to the black hole entropy,
and the corresponding quantity in the absence of backreaction;
compared with the latter case, the entropy density including backreaction is wider and overall smaller. In both cases, the entropy density is substantial in a region for which $h(\rho)$ is not parametrically small. Furthermore, with an appropriate choice of $\Phi_0$, this is also a weak coupling region.

\section{Discussion}

We argued that the backreaction of the winding mode in the $SL(2,\mathbb{R})_k/U(1)$ cigar leads, at least in the large $k$ limit, to a puncture at its tip, and that the winding mode carries the black hole entropy.  This raises two natural  questions. First, what does this imply for the Lorentian $SL(2,\mathbb{R})_k/U(1)$  black hole? Second, is this special for the $SL(2,\mathbb{R})_k/U(1)$ cigar? Or should we expect a similar phenomenon in, say, the Schwarzschild cigar?

\begin{figure}
	\centering
	\includegraphics[scale=0.5]{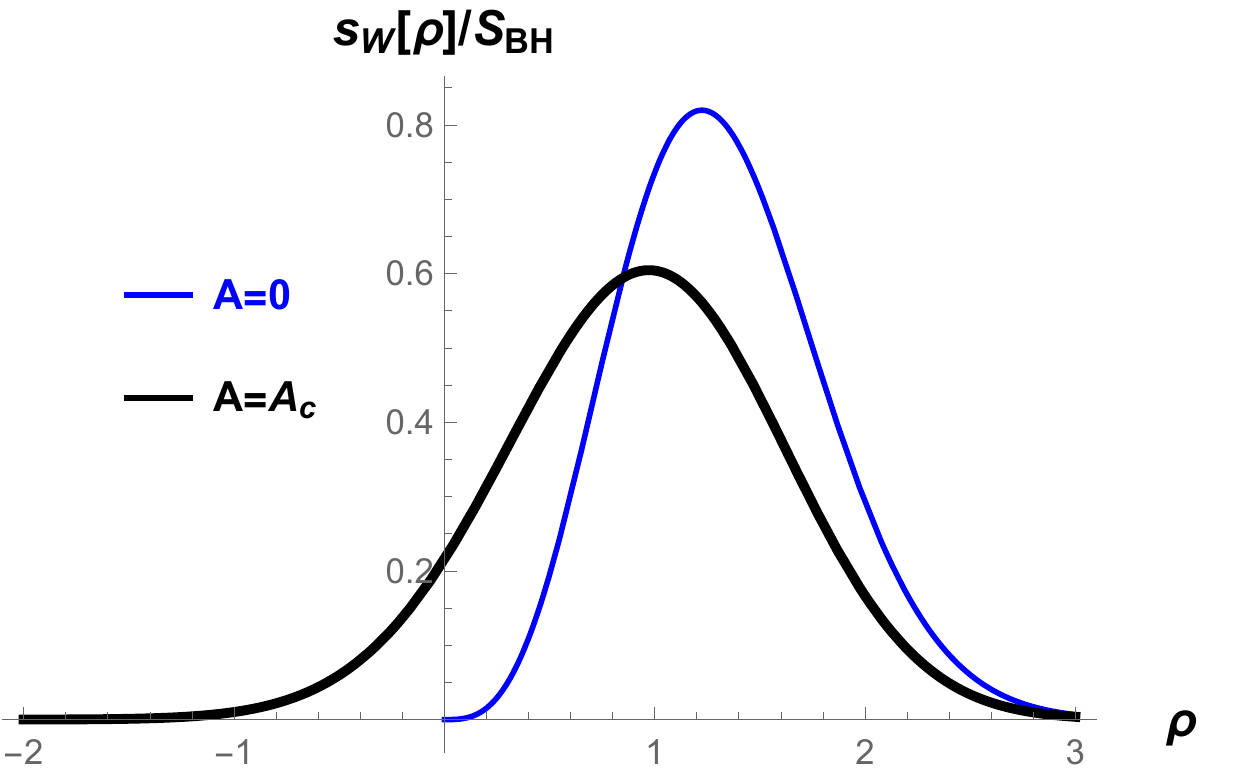}
	\caption{Shown is the ratio of the entropy density of the winding-mode condensate to the black hole entropy for the critical solution (thick, black), and for a solution neglecting the backreaction (thin, blue).}
\end{figure}

We start with the first question. The analytic continuation of a cigar with a puncture at its tip is a black hole without  a future wedge.  Heuristically, it is as if the singularity got expelled to the horizon.  The analytic continuation of the winding modes gives~\cite{Giveon:2019gfk,Giveon:2019twx} (see~\cite{Jafferis:2021ywg} for a related discussion)
an instant folded string~\cite{Itzhaki:2018glf} that fills the black-hole interior.
The energy-momentum tensor of an instant folded string is quite non-trivial~\cite{Attali:2018goq}
-- for example, it violates the averaged null energy condition. As a result, a sufficient number of them render the
$SL(2,\mathbb{R})_k/U(1)$ black hole impenetrable~\cite{Giveon:2020xxh}.
This, as well as the stringy glimpse into the black-hole interior studied in~\cite{Ben-Israel:2017zyi,Itzhaki:2018rld},
seems to indicate that, in a sense, the information is ejected from the black hole through the puncture.
Clearly, a more detailed study is required to make this claim precise.

Let us turn to the second question. In light of our findings here, it is tempting to conjecture that in string theory there is always a puncture at the tip of cigars (including the Schwarzschild cigar). Namely, that
\be\label{conj}
A_s=A_c~.
\ee
However, without an exact worldsheet CFT description it is hard to calculate both $A_s$ and $A_c$.~\footnote{Note that the exact CFT description of the $SL(2,\mathbb{R})_k/U(1)$ was crucial for justifying the use of (\ref{goku}) for a large black hole, and so it entered also into the calculation of  $A_c$.}

There is however an avenue~\cite{Giveon:2021gsc} to proceed:
Reducing the Schwarzschild metric on the sphere, $S^{D-2}$, one gets~\cite{Emparan:2013xia} in its near horizon an $SL(2,\mathbb{R})_k/U(1)$ black-hole background (times a sphere). There is a sense in which this effective description becomes an exact worldsheet CFT  of the near-horizon physics~\cite{Emparan:2013xia,Chen:2021emg,Giveon:2021gsc} (at least to sufficient approximation).
This might thus provide a path to explore~(\ref{conj}), as well as other aspects~\cite{Giveon:2021gsc} of large $D$-dimensional black holes in string theory.

\section*{Acknowledgments}

The work of AG and NI is supported in part by a center of excellence supported by the Israel Science
Foundation (grant number 2289/18) and BSF (grant number 2018068). YZ is supported in part by the Adams fellowship.

\vspace{10mm}

\appendix

\section{ Equations of motion}

In this appendix, we sketch some steps leading to the complete set of three equations of motion in~(\ref{bad}),
as well as the special equations in~(\ref{goku}), which appear in Sections 2 and 5.

\subsection{HP equations of motion}

The initial equations are the HP equations of motion,~\cite{Horowitz:1997jc}:
Varying the HP action in (\ref{HP}) w.r.t. the dilaton yields
\begin{eqnarray}\label{Dilaton}
&R-2\Lambda + 4 \nabla^2 \Phi - 4 g^{\mu \nu} \partial_{\mu} \Phi \partial_{\nu} \Phi
&= 2\kappa^2 \left(g^{\mu \nu}\partial_{\mu} \chi \partial_{\nu} \chi^* +\frac{\beta ^2 g_{\tau \tau}-\beta_H ^2}{(2\pi )^2}\chi \chi^*\right).
\end{eqnarray}
Combining this with the $g_{\tau \tau}$ metric equation of motion, it follows that
\begin{eqnarray}\label{tt}
&R_{\tau \tau} +2\nabla _{\tau} \nabla_{\tau} \Phi=
&-2\kappa^2  \frac{\beta ^2 g_{\tau \tau}^2}{(2\pi )^2}\chi \chi^*~.
\end{eqnarray}
Similarly, for the spatial metric equation of motion, one has
\begin{eqnarray}\label{rr}
&R_{\mu \nu}+2\nabla _{\mu} \nabla_{\nu} \Phi=
&2\kappa^2  \partial_{\mu} \chi \partial_{\nu} \chi ^*~.
\end{eqnarray}
Finally, the $\chi$ equation of motion reads
\begin{equation}
\frac{e^{2\Phi}}{\sqrt{g}} \partial_\mu\left(\sqrt{g} e^{-2\Phi } g^{\mu \nu}\partial_{\nu} \chi \right)=\frac{\beta ^2 g_{\tau \tau}-\beta_H ^2}{(2\pi )^2}\chi~.
\end{equation}

In the geometry (\ref{ansatz}), and with $\Phi=\Phi(\rho)$ and $\chi=\chi (\rho)$, we obtain:
\begin{eqnarray}\label{dilaton}
&h''(\rho)-\frac{2}{k} h(\rho)  -2h'(\rho)\Phi' (\rho) -2h(\rho)\Phi'' (\rho)+2h(\rho)(\Phi'(\rho))^2=\nonumber \\
&-h(\rho) \left[\chi'(\rho)\chi'(\rho)^*  +(h^2(\rho)-2)\chi (\rho)\chi^* (\rho)\right]~,
\end{eqnarray}
\begin{equation}\label{tautau}
h''(\rho) - 2 h' (\rho)\Phi'(\rho) = 2h^3(\rho) \chi (\rho)^2~,
\end{equation}
\begin{equation}\label{xx}
h''(\rho)-2h\Phi''(\rho) =- 2 h(\rho)\chi'(\rho)^2~,
\end{equation}
\begin{equation}\label{chi}
h(\rho)\chi ''(\rho)+h'(\rho) \chi '(\rho) -2 h(\rho)\chi ' (\rho) \Phi' (\rho) =(h^2(\rho) - 2) h(\rho) \chi(\rho)~.
\end{equation}
The difference between eqs. (\ref{tautau}) and (\ref{xx}) leads to:
\begin{equation}\label{comb1}
h\left(\frac{\Phi'}{h}\right)' = (\chi')^2+h^2\chi^2~.
\end{equation}
A substitution of (\ref{tautau}) into  (\ref{dilaton}) implies
\begin{equation}\label{comb2}
\frac{2}{k}+2\Phi''-2(\Phi')^2= (\chi')^2 + (3h^2-2)\chi^2~.
\end{equation}
Equations (\ref{chi}), (\ref{comb1}) and (\ref{comb2}) are the three equations in~(\ref{bad}).

Next, we would like to derive  (\ref{tautau}) from the equations in (\ref{bad}), namely, from eqs. (\ref{chi}),(\ref{comb1}),(\ref{comb2}):
Multiplying  (\ref{comb1}) by $-h$, and then taking a derivative, lead to the following equation,
\begin{equation}
\label{comb3}
h'' \Phi'-h\Phi'''=-h' (\chi')^2 -2h \chi' \chi'' -3h^2 h' \chi^2 -2h^3 \chi \chi'~.
\end{equation}
The derivative of  (\ref{comb2}) is given by
\begin{equation}
\label{comb4}
\Phi'''-2\Phi'' \Phi' = \chi'' \chi' + (3h^2-2) \chi \chi' +3hh' \chi^2~.
\end{equation}
Combining eqs.~(\ref{comb3}), (\ref{comb4}) to eliminate $\Phi'''$, results in yet another equation,
\begin{equation}
\label{Comb3}
h'' \Phi'-2h\Phi''\Phi'=-h' (\chi')^2 -h \chi' \chi'' +h (h^2-2)\chi \chi'~.
\end{equation}
Equation~(\ref{chi}) multiplied by $\chi'$ gives:
\begin{equation}
\label{comb5}
h \chi' \chi''+h' (\chi')^2 = 2h (\chi')^2 \Phi' + h(h^2-2)\chi \chi'~.
\end{equation}
Substituting (\ref{comb5}) into (\ref{Comb3}), one obtains the following equation,
\begin{equation}
h''\Phi' -2h \Phi'' \Phi' = -2h (\chi')^2 \Phi'~.
\end{equation}
Dividing by $\Phi'$ and using (\ref{comb1}), we thus obtain (\ref{tautau}).

To recapitulate, we verified that the three equations in (\ref{bad}) provide a complete set of independent equations of motion
for the ansatz (\ref{ansatz}) for the metric, with a dilaton and winding mode $\Phi=\Phi(\rho)$ and $\chi=\chi (\rho)$, respectively.

\subsection{Simplified (\ref{goku})}

In this Subsection, it is demonstrated that any solution of the equations in~(\ref{goku}),
\begin{equation}\label{Phi'}
h'=h\Phi'+1~,
\end{equation}
\begin{equation}\label{chi^2}
\Phi'=-h\left(\chi^2+\frac{1}{k}\right),
\end{equation}
\begin{equation}\label{chi'}
\chi ' = -h \chi~,
\end{equation}
also solves (\ref{bad}).

First, using eqs.~(\ref{chi'}) and (\ref{Phi'}), the L.H.S. of the second equation of (\ref{bad}), namely (\ref{chi}),
can be shown to be equal to its R.H.S.,
\begin{equation}
h \chi '' + h' \chi ' - 2 h \chi' \Phi' = h(h^2-2)\chi~.
\end{equation}
Second, the derivative of  (\ref{chi^2}) is
\begin{equation}\label{ABad}
\left(\frac{\Phi'}{h}\right)'=-2\chi \chi'~.
\end{equation}
Then (\ref{chi'})  implies
\begin{equation}
h\left(\frac{\Phi'}{h}\right)'=(\chi')^2+h^2\chi^2~.
\end{equation}
Finally, we derive
\begin{equation}
\frac{2}{k}+2\Phi '' - 2(\Phi')^2 =(\chi')^2+ (3h^2-2)\chi^2~.
\end{equation}
Eq. (\ref{ABad}) can be rewritten as:
\begin{equation}
\Phi '' - \frac{h'}{h} \Phi' = 2h^2 \chi^2~.
\end{equation}
Eq. (\ref{Phi'}) implies that
\begin{equation}
\Phi '' -\left(\Phi'+\frac{1}{h}\right) \Phi' = 2h^2 \chi^2~.
\end{equation}
Thus,
\begin{equation}
\Phi '' - (\Phi')^2 = 2h^2 \chi^2+\frac{\Phi'}{h}~.
\end{equation}
Eq. (\ref{chi^2}) implies
\begin{equation}\label{conc}
\frac{1}{k}+\Phi '' - (\Phi')^2 = (2h^2-1) \chi^2=\frac{1}{2}(\chi')^2+ \left(\frac{3}{2}h^2-1\right)\chi^2~.
\end{equation}
This establishes that by solving  (\ref{goku}), one obtains a solution of (\ref{bad}).

\section{  $A_s=\exp{(-\gamma/2)}$}

In sections 2, 3 and 5, we showed that there is a critical value of the amplitude, $A$,
of the winding condensate $\chi$ (see~(\ref{coshk},\ref{cfzz},\ref{critical})),
for which its backreaction on the cigar background, in the parametrically small curvature limit,
develops a puncture at its origin, such that, in particular, the thermal circle does not shrink at finite radial direction,
and the entropy that it then carries is precisely that of the Bekenstein-Hawking one.

In this appendix, we show that the critical value of $A$ in sections 2 and 3, $A_c$ in (\ref{critical}), corresponds precisely
to the coupling of the winding condensate in the large $k$ limit of the exact $SL(2,\mathbb{R})_k/U(1)$ cigar SCFT, $A_{s}$ in~(\ref{cfzz}).
Namely, in string theory on the Euclidean black hole,
{\it the whole entropy is carried by the winding string at the horizon}.

The material used in the following relies heavily on the manipulations done long ago and more recently in a set of papers;
in particular,~\cite{Giveon:2019twx} and references therein include an essential piece of the required material.
As discussed there, consider the worldsheet (WS) action on the cylinder, $R_\rho\times S^1_\tau$, $\tau\sim\tau+2\pi\sqrt{k}$,
with a linear dilaton,
\be\label{flatld}
S_0=\frac{1}{2\pi}\int d^2z\left(\partial\rho\bar\partial\rho+\partial\tau\bar\partial\tau-\frac{1}{\sqrt k}\hat{R}\rho\right)~,
\ee
where $\rho$ and $\tau$ are canonically normalized fields.~\footnote{Rcall that we chose $\alpha'=1$.}
Since we consider the type II string theory, the action has additional terms, which include worldsheet fermions,
such that it has an $N=(2,2)$ superconformal symmetry; these are not essential for us, and we thus ignore them here and below.

The $SL(2,\mathbb{R})_k/U(1)$ SCFT is obtained, e.g., by turning on the graviton and $N=2$ Liouville interactions,
\be\label{sint}
S_{int}=\frac{1}{2\pi}\int d^2z\left(2\pi\frac{k}{2}\lambda_I^{WS}\partial\tau\bar\partial\tau e^{-2\rho/\sqrt{k}}
+2\pi\frac{1}{2}\lambda_W^{WS}\frac{1}{k}\left(\int d^2\theta  e^{-\sqrt{k}\Psi}+c.c.\right)\right)~,
\ee
respectively,
where $\Psi=\rho+i\tilde\tau+...$, with $\tilde\tau\equiv\tau_L-\tau_R$ being the T-dual of the thermal circle $\tau\equiv\tau_L+\tau_R$,
is a chiral superfield (the dots stand for its higher components).~\footnote{See, e.g,.~\cite{Nakayama:2020tag} and references therein
for the properties of the $N=2$ Liouville term in (\ref{sint}).}
Finally, the coupling of the winding condensate, $\lambda_W^{WS}$, is related to that of the graviton, $\lambda_I^{WS}$, via~\cite{Giveon:2001up}
\be\label{lambdaw}
\lambda_W^{WS}=\frac{k}{\pi}\left(2\pi\lambda_I^{WS}\frac{k}{2}\frac{\Gamma(1+1/k)}{\Gamma(1-1/k)}\right)^{k/2}~.
\ee
This property of the exact SCFT is an essential ingredient the discussion below.~\footnote{The $\frac{1}{2\pi}\int d^2z\left(2\pi...\right)$
in (\ref{sint}) is emphasizing the $2\pi$ factors
that will appear below: the $\frac{1}{2\pi}$ is the factor required in the kinetic term (\ref{flatld})
for canonically normalized fields (see, e.g., section 2.1 in~\cite{Polchinski:1998rq}) and the $2\pi$ in the integrand,
as well as other factors, are for the $\lambda's$ in (\ref{sint}) to be
normalized as those in (\ref{lambdaw}); concretely, in  (2) of~\cite{Bershadsky:1991in},
there is a $\frac{1}{2\pi}\int 2\pi\mu\beta\bar\beta...=\frac{1}{2\pi}\int 2\pi\mu\frac{k}{2}\partial\tau\bar\partial\tau...$,
and $\lambda_I^{WS}\equiv\mu$ (this follows from the relation $\beta=\sqrt{k}\partial(\rho+i\tau)$, as can be seen in~\cite{Giveon:2019twx},
following (17)--(24) in~\cite{Bershadsky:1991in}).}

From~\cite{Teschner:1999ug,Giveon:1999px,Giveon:2001up,Giveon:2015cma},
we know that the stringy contribution to the
reflection coefficient, for scattering from the tip of the cigar CFT, is
\be\label{deltanp}
R_{s}(j)\equiv e^{-i\delta_{stringy}(j)}=\frac{\Gamma(1-(2j+1)/k)}{\Gamma(1+(2j+1)/k)}~,
\ee
where $j$ is related to the radial momentum in the cigar, as described, e.g., in~\cite{Giveon:2015cma}.
Hence,
\be\label{list}
\sqrt{R_s(0)}\lambda_I^{ST}=2\pi\lambda_I^{WS}\frac{k}{2}~,
\ee
where $\lambda_I^{ST}$ is the coefficient of $e^{-2\rho/\sqrt{k}}$ in the asymptotic behavior of $g_{\tau\tau}=h^2$ in spacetime (ST),
and consequently
\be\label{lwst}
\lambda_W^{WS}=\frac{k}{\pi}\left(\lambda_I^{ST}\sqrt\frac{\Gamma(1+1/k)}{\Gamma(1-1/k)}\right)^{k/2}~.
\ee
The $2\pi$ in (\ref{list}) arises when going from the WS to ST due to the $2\pi$ range of the string parameter
$\sigma_1$ in the $z=e^{\sigma_2+i\sigma_1}$ plane~\footnote{Technically, it is due to the $2\pi$ in the integrand of (\ref{sint}).}
and the $k/2$ is due to the periodicity in the $\tau$ direction, both of which are explicitly reflected in (\ref{sint}).
Finally, the $\sqrt{R_s(0)}$ on the left hand side of (\ref{list})
is a stringy dressing of the graviton one point function in spacetime,
required for the appropriate map from the graviton condensate~\footnote{Whose imaginary radial momentum amounts to $j=0$
in (\ref{deltanp}).}
in the exact worldsheet CFT, to the ST geometry variables~\footnote{Which
are a priori blind to this non-perturbative (in $\alpha'$), strictly stringy correction, in the spacetime effective theory.}.

The second equation in~(\ref{twokc}) gives
\be\label{lambdaist}
\lambda_I^{ST}=4~,
\ee
and $\lambda_W^{ST}$ is the coefficient of $e^{-\sqrt{k}\rho}$ in the asymptotic behavior of the winding condensate field $\chi$
in spacetime, namely (see the first equation in~(\ref{twokc})),
\be\label{lambdawst}
\lambda_W^{ST}=2^kA_s~.
\ee
The second term in (\ref{sint}) gives
\be\label{relation}
\lambda_W^{ST}=\frac{\pi}{k}\lambda_W^{WS}~,
\ee
and thus, plugging (\ref{lwst}) with (\ref{lambdaist}) in (\ref{relation}), and using~(\ref{lambdawst}), in the large $k$ limit we find
\be\label{largek}
A_s=e^{-\gamma/2}~,
\ee
where $\gamma$ is the Euler-Mascheroni constant.
We have thus derived (\ref{cfzz}).

\section{Critical behavior}
Here, we quantify the properties of the solutions as the amplitude approaches the critical value from below. The solutions exhibit a universal critical behaviour (see Figs. 6 and 7): We find that the location of the tip $\rho_{tip}$ and of $\rho_{min}$, the location of the minimum  of $h'$,  are pushed to negative values of $\rho$. To a good approximation, we find that the critical exponents in $\rho_{tip}\propto (A_c-A)^{-\gamma_t}$, $\rho_{min}\propto (A_c-A)^{-\gamma_m}$ are $\gamma_t \approx 0.05~,~ \gamma_m \approx 0.09$.
\begin{figure}[h]
	\centering
	\includegraphics[scale=0.6]{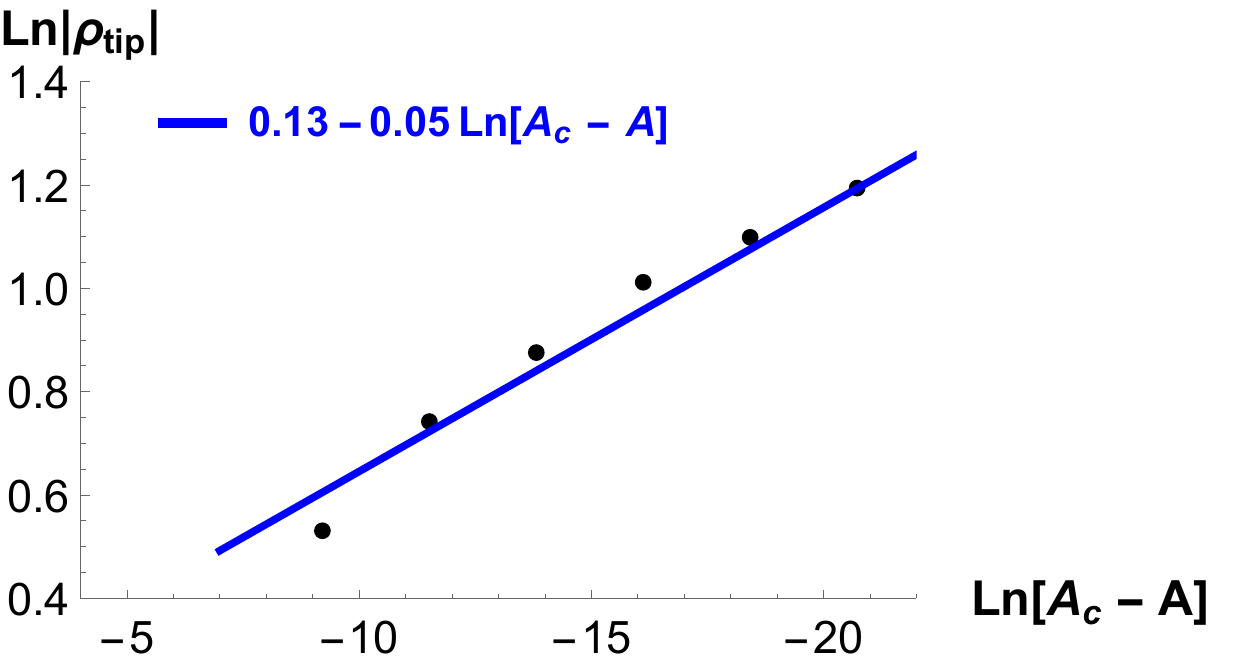}
	\caption{Depicted is the logarithm of $-\rho_{tip}$ as a function of the logarithm of the deviation from the critical amplitude. The slope of a linear fit is found to be $  -0.05\pm 0.01$.}
\end{figure}

\begin{figure}[h]
	\centering
	\includegraphics[scale=0.6]{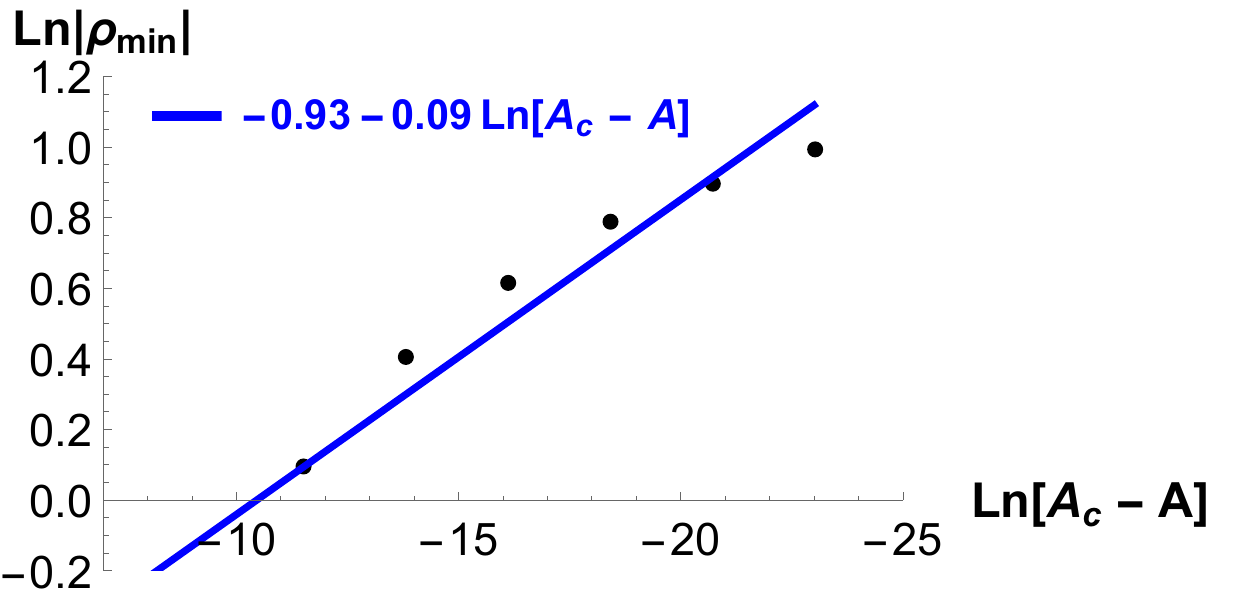}
	\caption{Depicted is the logarithm of $-\rho_{min}$ as a function of the logarithm of the deviation from the critical amplitude. The slope of a linear fit is found to be $  -0.09\pm 0.025$.}
\end{figure}

\end{document}